\documentclass{svjour3}                     

\smartqed  
\usepackage{graphicx}
\usepackage{amsmath}
\usepackage{color}
\usepackage{amssymb}
\RequirePackage{natbib}
\usepackage{url}
\RequirePackage{ifthen}
\newcommand{\href}[2]{{#2}}
\usepackage{mathptmx}      

\newcommand{\rfia}[1]{\makebox[\parindent][l]{%
                     \makebox[0em][r]{\rm(}\sf#1\rm)}}
\newcounter{ABCcB}
\newcommand{\theABCcC}{\alph{ABCcB}}

\newenvironment{ABCTH}{\begin{list}{
  \rfia{\theABCcC}}{\usecounter{ABCcB} \topsep 0ex \partopsep 0ex \itemsep0ex
  \parsep=\parskip \leftmargin 0em \rightmargin 0em \itemindent=\parindent
  \listparindent=\parindent  \labelsep 0.5em \labelwidth 0.5em }}{\end{list}}

\newcommand{\Ew}{\mathop{\rm {{}E{}}}\nolimits} 
\newcommand{\ve}{\varepsilon}

\newcommand{\ssr}{\rm\scriptscriptstyle}
\newcommand{\Lo}{\mathop{\rm {{}o{}}}\nolimits}
\newcommand{\LO}{\mathop{\rm {{}O{}}}\nolimits}

\newcommand{\Jc}{\mathop{\rm I}\nolimits}

\newcommand{\R}{\mathbb R}

\newcommand{\iid}{\mathrel{\stackrel{\ssr i.i.d.}{\sim}}}

\newtheorem{Thm}{Theorem}
\newtheorem{Prop}[Thm]{Proposition}
\newtheorem{Lem}[Thm]{Lemma}
\newtheorem{Rem}[Thm]{Remark}

\newtheorem{Def}[Thm]{Definition}

\renewcommand{\eqref}[1]{(\ref{#1})}

\numberwithin{equation}{section}
\numberwithin{Thm}{section}
\begin{document}
\ifx\blinded\undefined
\title{Yet another breakdown point notion: EFSBP\newline
{\large---illustrated at scale-shape models\thanks{This work was supported
by a DAAD scholarship for N.H.}}
}

\author{ Peter Ruckdeschel        \and
         Nataliya Horbenko 
}

\institute{P.~Ruckdeschel \and N.~Horbenko \at
              Fraunhofer ITWM, Department of Financial Mathematics, \\
              Fraunhofer-Platz 1, D-67663 Kaiserslautern\\
              and Dept.\ of Mathematics, University of Kaiserslautern,\\
              P.O.Box 3049, D-67653 Kaiserslautern \\
              \email{peter.ruckdeschel@itwm.fraunhofer.de\\           
              \hphantom{E-mail: }nataliya.horbenko@itwm.fraunhofer.de}
}
\else
\title{Yet another breakdown point notion: EFSBP\newline
{\large---illustrated at scale-shape models}}
\author{}
\institute{}
\fi
\date{Received: date / Accepted: date}

\maketitle
\begin{abstract}
The breakdown point in its different variants is one of the
central notions to quantify
the global robustness of a procedure. 
%
We propose a simple supplementary variant
which is useful in situations where we have no obvious or only partial
equivariance: Extending the \citet{Do:Hu:83} \textit{Finite Sample Breakdown Point \/}, we propose the
\textit{Expected Finite Sample Breakdown Point \/} to
produce less configuration-dependent values while still preserving
the finite sample aspect of the former definition.

We apply this notion for
joint estimation of scale and shape (with only scale-equivariance available),
exemplified for generalized Pareto, 
generalized extreme value,
Weibull, and Gamma distributions.

In these settings, we are interested in highly-robust,
easy-to-compute initial estimators; to this end we study
Pickands-type and Location-Dispersion-type estimators
and compute their respective breakdown points.

\end{abstract}
\section{Introduction}
In an industrial project to compute robust variants of \textit{OpVar},
i.e.; the regulatory capital as required in \citet{ICCMCS:04}
for a bank to cover its \textit{operational risk}, we came across the
problem of determining the (finite sample) breakdown point of
certain considered procedures. Here operational risk is by definition
``the risk of direct or indirect loss resulting from inadequate or
failed internal processes, people and systems or from external events.''

These extremal events, as motivated by the Pickands-Balkema-de\,Haan
Extreme Value Theorem (see \citet{B:H:74}, \citet{Pick:75})
suggest the use of the generalized Pareto distribution (GPD) for modeling in this context.
In an intermediate step this modeling involves estimation of the
scale and shape parameters of this distribution.
To this end, several robust procedures have been proposed
in the literature%
\ifx\blinded\undefined
, see \citet{Ru:Ho:10} for a more detailed discussion.
\else
.
\fi

One of the quantities to judge robustness of a procedure is the
breakdown point (see Definition~\ref{BPdef}). In particular, we are interested
in the finite sample version FSBP of this notion to be able to quantify
the degree of protection a procedure provides in the estimation at
an actual (finite) set of observations.

It turns out that for our purposes the original definition has some drawbacks,
as it depends strongly on the configuration of the actual sample.
To get rid of the dependence on possibly highly improbable sample
configurations while still preserving the aspect of a finite sample, we
propose an expected FSBP, \textit{EFSBP}, i.e.; to integrate out the FSBP with
respect to the ideal distribution.

We illustrate the usefulness of this new concept for scale-shape models
by means of two types of robust estimators, quantile-type estimators
(Pickands Estimator \textit{PE})
and robust Location-Dispersion
(LD) estimators as introduced by \citet{Ma:99}; for the latter type we study estimators based on the median
for the location part and several robust scale estimators for the dispersion part:
a (new) asymmetric version of the median of absolute deviations kMAD,
as well as Qn and Sn from \cite{Ro:93}---combined to \textit{MedkMAD}, \textit{MedQn}, and \textit{MedSn},
respectively.

These estimators are meant to be used as initial estimators
with acceptable to good global robustness properties for (more efficient) robust estimators afterwards.
In particular, they can be computed \textit{without} the need of additional (robust, consistent) initial
estimators, which precludes otherwise promising alternatives like Minimum Distance estimators, for
which we could have read off asymptotic breakdown point values as high as half the optimal value
from \cite{Do:Li:88}.  We have also excluded the method-of-median approach of \cite{P:W:01}, because
in contrast to PE and MedkMAD, MedQn, and MedSn, for this estimator
in the GPD and GEVD case, no explicit calculations are possible. We have studied this approach in another
paper, though (\cite{Ru:Ho:10}), and empirically found that in the GPD case its breakdown behavior
is worse than the one of MedkMAD and MedQn.

Our paper is organized as follows: In Section~\ref{Model Setting}, we list our reference examples for
scale-shape models, i.e.; the generalized
Pareto, the generalized extreme value, the Weibull,
and the Gamma distribution, as well as the Gross Error model which we use to
capture deviations from the ideal model. In
Section~\ref{Robustness}, we recall  the standard definitions of the asymptotic
and finite sample breakdown
points ABP and FSBP and introduce the new concept of EFSBP in Definition~\ref{AFSBPDef}.
Section~\ref{EstDef} then defines the considered estimators, i.e.; quantile-type estimators PE,
and 
LD estimators MedkMAD, MedQn, MedSn. At these estimators, we
demonstrate our new breakdown point notion in Section~\ref{EstBP},
giving analytic formulae for FSBP, ABP, and EFSBP in
Propositions~\ref{PEprop}, \ref{MedkMADprop},
and \ref{Nnprop}, together with some numerical evaluations of EFSBP at some
reference situation and with simulation-based evaluations.
Proofs for our results are gathered in Appendix~\ref{proofsec}.
\ifx\blinded\undefined
\begin{Rem}
\rm\small This paper is a part of the PhD thesis of the second author;
a preliminary version of it may be found in Ruckdeschel and Horbenko (2010).
\end{Rem}
\fi
\section{Model Setting}\label{Model Setting}
%
For notions of invariance of statistical models and equivariance of estimators we refer to \cite{Ea:89}:
Given a measurable space $(\Omega,\mathcal{B})$, a family of probability measures $\mathcal{P}$ defined on $\mathcal{B}$ is a \textit{statistical model}.\\
{\small Notationally, we use the same symbol for the cumulative distribution function (c.d.f.) and the probability measure;
we write $F(x-{\scriptstyle 0})$ to denote left and, correspondingly, $+{\scriptstyle 0}$ for right limits, and $F^{-}$ to denote the right continuous quantile function
given by $F^-(s)=\inf\{t\in\R\colon\;F(t)\ge s\}$.}
\begin{definition}
Suppose a group $G$ acts measurably on $\Omega$. Model $\mathcal{P}$ is called \textit{$G$-invariant} iff for each $P\in \mathcal{P}$, the image
probability $gP$ of $P$ under group action $g$ stays in  $\mathcal{P}$.
\end{definition}
For simplicity, we assume that $g(P_1)=g(P_2)$ implies $P_1=P_2$ for any two elements of $\mathcal{P}$.
In a $G$-invariant parametric model $\mathcal{P} = \{P_\theta|\theta \in \Theta\}$, where $\Theta$ is the parameter space,
group $G$ induces an isomorphic group $\tilde G$, acting on the parameter
space with the identification $g(P_\theta)=P_{\tilde g(\theta)}$.
 In this situation, a point estimator $t$ mapping $\Omega$ to $\Theta$ is equivariant iff
$$t(g(x)) = \tilde g(t(x)).$$
\subsection{Generalized Pareto Distribution and Other Scale-Shape Families}\label{GPDsec}
We illustrate our concepts at scale-shape models; our reference example is the three-parameter
generalized Pareto distribution (GPD) which has
c.d.f.\ and density
\begin{eqnarray}
&&F_{\theta}(x) = 1- \left( 1 + \xi \frac{x - \mu}{\beta} \right)^{-\frac{1}{\xi}},
\quad f_{\theta}(x) = \frac{1}{\beta}
\left( 1 + \xi \frac{x - \mu}{\beta} \right)^{-\frac{1}{\xi} - 1}
\end{eqnarray}
where $x \geq \mu$ for $\xi \geq 0$, and $\mu < x \leq \mu - \frac{\beta}{\xi}$
if $\xi < 0$. It has parameter $\theta=(\xi,\beta,\mu)^\tau$, for location
$\mu$, scale $\beta>0$ and shape $\xi$. Special cases of GPDs are the uniform
($\xi = -1$), the exponential ($\xi = 0$, $\mu=0$), and Pareto ($\xi > 0$,
$\beta=1$) distributions. We limit ourselves to the case of known
location $\mu=0$ and unknown scale and shape
here and abbreviate the pair $(\beta,\xi)$ by $\vartheta$, i.e.;
we are concerned with joint estimation of $\vartheta=(\beta,\xi)$ only.

Other scale-shape families for which our considerations apply mutatis mutandis
are the \textit{generalized extreme value distribution} (GEVD) given by its c.d.f.\
\begin{equation}
F_{\theta}(x)=\exp\Big( - \left( 1 + \xi \frac{x - \mu}{\beta} \right)^{-\frac{1}{\xi}}\Big)\Jc_{(-\frac{\beta}{\xi}+\mu,\infty)}(x)
\end{equation}
the \textit{Weibull distribution} with density
\begin{equation}
f_{\vartheta}(x)=\frac{\xi}{\beta}\Big(\frac{x}{\beta}\Big)^{\xi-1}\exp( - (x/\beta)^\xi) \Jc_{(0,\infty)}(x)
\end{equation}
and the \textit{Gamma distribution} with density
\begin{equation}
f_{\vartheta}(x)=\frac{x^{\xi-1}}{\beta^\xi\Gamma(\xi)}\exp( - (x/\beta)) \Jc_{(0,\infty)}(x)
\end{equation}
For the Weibull and Gamma case we require $\xi>0$, whereas in the GEVD case the same distinction
applies as in the GPD case.

\paragraph{Reparametrization}
In the Weibull family, passage to the log-observations transforms this model into
a location-scale model with the standard Gumbel as central distribution.
This approach has been taken by \citet{B:C:C:09}, and allows them to recur to the rich
theory (both classical and robust) available for location-scale models.

In both GPD and GEVD, a similar approach is possible, once instead of $\mu$ we use
$\tilde \mu = \mu\xi-\beta$, so that in this setting we get
\begin{equation}
1+ \xi \frac{x - \mu}{\beta} = \xi \frac{x - \tilde \mu}{\beta}
\end{equation}
In the GPD case, this leads to a location-scale model with the standard Exponential as central
distribution. This parametrization is used for two-parameter Pareto distribution, e.g.\ in \citet{Br:Se:00}. Two issues, however, are bought with this approach: First, knowledge of $\mu$
is not the same as knowledge of $\tilde \mu$, so our original setting where $\mu$ was assumed
known does not carry over easily. Second, the corresponding transformed model about the Exponential
distribution is not smooth---$L_2$-differentiable to be precise. The reason for this is essentially
that observations around the left endpoint of the distribution carry overwhelmingly much information
about the location parameter. As a consequence, usual optimality theory no longer is available,
and in the ideal model setting there are estimators which are consistent at faster rates than the
usual $1/\sqrt{n}$. On the other side, this high accuracy requires to base inference essentially
completely on the minimal observations which makes these procedures extremely prone to outliers.
Robustifications avoid this problem, but still, due to the lack of smoothness no optimality
theory is available. For this reason, we stick to the original parametrization.

\paragraph{Our reference model}
In the sequel,  we use the reference values $\beta=1$ and $\xi=0.7$ for all our scale-shape
models; in case of the GPD this amounts to moderately fat tails which reflects well
the situation we met in our application to OpVar.
%
%

\paragraph{In-/equivariance} The reduced model enjoys a certain \textit{invariance}:
with an included scale component, it remains invariant under scale
transformations $s_\beta(x)=\beta x$ of the observations.
Using the matrix $d_\beta=\mathop{\rm diag}(\beta,1)$, this invariance
is reflected by a corresponding notion of equivariance of estimators,
i.e.; an estimator $S$ for $\vartheta=(\beta,\xi)$ is called
\textit{scale-equivariant} if
\begin{equation}\label{scaleeq}
S(\beta x_1,\ldots, \beta x_n)= d_\beta S(x_1,\ldots, x_n)
\end{equation}

For the shape parameter $\xi$, there is no obvious such invariance,
entailing a dependence of estimator properties like robustness
on this parameter.


\subsection{Gross Error Model}
Extending the ideal model setting, Robust Statistics defines
suitable distributional neighborhoods about this ideal model. In this
paper, we limit ourselves to the \textit{Gross Error Model}, i.e.;
as neighborhoods, we use the sets of all distributions $F^{\ssr re}$
representable as
\begin{equation}\label{GEM}
F^{\ssr re}=(1-\ve) F^{\ssr id} + \ve F^{\ssr di}
\end{equation}
for some given size or radius $\ve>0$,
where $F^{\ssr id}$ is the underlying ideal distribution and $F^{\ssr di}$ some
arbitrary, unknown, and uncontrollable contaminating distribution.
\section{Global Robustness: the Breakdown Point}\label{Robustness}
In this paper we focus on the \textit{Breakdown Point}
as a global measure of robustness, specifying the reliability
of a procedure under massive deviations from the ideal model.
In the gross error model~\eqref{GEM}, it gives the largest radius $\ve$
at which the estimator still produces meaningful results.

In standard literature on Robust Statistics, there are two notions
of breakdown point---the \textit{asymptotic {\textrm(functional)}
breakdown point} (\textit{ABP})  and
the \textit{finite sample breakdown point} (\textit{FSBP})
introduced in \citet{Ha:68} and \cite{Do:Hu:83}, respectively:

\begin{Def} \label{BPdef}
\begin{ABCTH}
\item \label{def:asy_breakdown_point} \citep[2.2~Definition~1]{Ha:Ro:86}
The \emph{asymptotic breakdown point (ABP)}  $\ve^\ast$ of the sequence of
estimators $T_n$ for parameter $\theta\in\Theta$ at probability $F$ is given by
\begin{eqnarray}
\ve^\ast:=\sup\Big\{\ve \in (0,1]; &&\mbox{there is a compact set
$K_\ve \subset \Theta$ s.t.}\nonumber\\
&& \pi(F,G) < \ve\;\;\Longrightarrow\;\; G(\{T_n \in K_\ve\})
\stackrel{n\to\infty}{\longrightarrow} 1\,\Big\}
\end{eqnarray}
where $\pi$ is Prokhorov distance.
\item \label{def:fin_breakdown_point} \citep[2.2~Definition~2]{Ha:Ro:86}
The \emph{finite sample breakdown point (FSBP)}  $\ve_n^\ast$ of the
estimator $T_n$ at the sample $(x_1,...,x_n)$ is given by
\begin{equation} \label{FSBPdefH}
\ve_n^\ast(T_n; x_1,...,x_n) := \frac{1}{n} \max \Big\{ m;
\max_{i_1,...,i_m} \sup_{y_1,...,y_m}|T_n(z_1,...,z_n)| < \infty \Big\},
\end{equation}
where the sample $(z_1,...,z_n)$ is obtained by replacing the 
data points $x_{i_1},...,x_{i_m}$ by arbitrary values $y_1,...,y_m$.
%
\end{ABCTH}
\end{Def}
Note that $\ve_n^\ast$ from  \eqref{FSBPdefH} is by $1/n$ smaller than the \cite{Do:Hu:83} FSBP.\\
Definition~\ref{def:fin_breakdown_point} does not cover the scale case, where we must take
into account the possibility of implosion as well: As noted by an anonymous referee, otherwise
one could achieve arbitrarily high breakdown points by choosing estimators based on
two very low quantiles, which of course would not be stable at all---an argument
valid in the location-scale case as well.
A remedy for the scale parameter is given by the log-transformation as mentioned in
\cite{He:05}, i.e.;

\begin{equation} \label{BPdef2}
\ve_n^\ast(T_n; x_1,...,x_n) := \frac{1}{n} \max \Big\{ m;
\max_{i_1,...,i_m} \sup_{y_1,...,y_m}|\log(T_n(z_1,...,z_n))| < \infty \Big\},
\end{equation}
%
%

\paragraph{Breakdown and partial invariance}

By arguments given in \citet{D:G:05}, a certain equivariance of the considered
estimator under a suitable group of transformations is required to
obtain meaningful upper bounds for the breakdown point. In our scale-shape
models, however, as indicated in Section~\ref{GPDsec}, we canonically
only have scale invariance. This lack of complete equivariance does not
invalidate the cited authors' considerations, but rather these can be extended
to also cover this partial invariance:

While due to the lack of shape-equivariance, we conjecture that similar defective
 constructions, which produce breakdown points arbitrarily close to $1$ in
 the AR(1) case (as mentioned in \citet{Ge:Lu:05}), should be feasible in the
 pure shape case as well, in the joint scale-shape case, imposing scale-equivariance,
 we do obtain sensible upper bounds as such constructions are eliminated
 by this (partial) equivariance.

In particular, as the scale model is a submodel of our scale-shape model,
the corresponding upper bounds for the maximal breakdown point among
all scale-equivariant estimators from \citet[Thms.~3.1,3.2]{D:G:05}
remain valid in our setting without change.
Hence, in the sequel, we restrict ourselves to scale-equivariant
estimators.
%
In particular, following \citet[sec.~4.2]{D:G:07}, we note that with $n_0$ being
the highest frequency of a single data point
in the original sample,
\begin{equation} \label{n0n1}
\ve_n^\ast \leq \lfloor \frac{(n-n_0-1)_+}{2}\rfloor/n
\end{equation}
(adapted to \eqref{FSBPdefH}) among all scale-equivariant estimators.

\paragraph{Breakdown and restricted parameter space}

In the GPD and GEVD families, there are two canonical parameter spaces
for $\xi$: Either one does not impose any restriction, i.e.;
$\xi\in\R$---which could be seen as
``natural'' there, or one restricts $\xi$ to be positive (which
is the only possibility for the Weibull and Gamma case).

In the GPD and GEVD case, $\xi=0$  is a discontinuity as to the
statistical properties of the model, comparable to parameter values $\pm 1$
in the AR(1) model. While GPD and GEVD for $\xi<0$ have compact support,
in the AR(1) model $\pm 1$ mark the border of stationarity.
In both cases, the discontinuity only becomes visible when passing to
sequences of observations, in our case when motivating GPD and GEVD
by asymptotic arguments, i.e.; by the Pickands--Balkema-de\,Haan and Fisher-Tippet-Gnedenko
Extreme Value Theorems. To this end we need a uniformity over sets of quantiles
which gets lost when passing over the value $\xi=0$.
In particular, shape in the
GPD and GEVD models decides to which domain of attraction belongs the underlying distribution
in the corresponding Extreme Value Limit Theorems.
In both the scale-shape and the AR(1) case, it is hence well debatable to restrict
the parameter space accordingly, see \citet{Ge:Lu:05} and the rejoinder
in \citet[p.~1033]{D:G:05}.
E.g.; we are mainly interested in the case when $\xi > 0$, which corresponds
to heavy-tailed GPD / GEVD, and an estimate $\xi\le 0$ would lead to drastic under-estimation
of the corresponding operational risk.

%

In the sequel, for the GPD and GEVD cases, we hence consider both situations: with
and without restriction on the parameter space, i.e.; that $\xi>0$ or $\xi \in\R$.

Similar arguments could be carried out in case of shape estimation in
the Weibull case, where $0<\xi<1$ corresponds to heavy-tailed, $\xi \geq 1$
to light-tailed distributions; we do not pursue this further here.




\paragraph{Breakdown and finite samples}

As for our purposes, reliability at finite samples is of
primary interest, we will focus on the FSBP.

For deciding upon which procedure to take \textit{before} having made
observations, in particular for ranking procedures in a simulation study, the
FSBP from Definition~\ref{def:fin_breakdown_point} has some drawbacks:
It is deliberately probability-free and based on an actual sample
$(x_1,...,x_n)$, which we assume from the ideal situation for the moment.
Hence its value depends on the configuration of this sample.
This is desirable when checking safety of a procedure at an
actual data set, but also entails that for the estimators considered in this
paper, a generally valid value for FSBP does not exist, and the only possible
universal lower bound will be the minimal possible value of $0$;
and even if we made a sample-wise restriction, banning such samples
from the application of the estimator, we would have other ones to come
up with an FSBP of $1/n$ and so forth. This does not reflect
the situation to be expected in the ideal model, though.
Hence, we follow the general spirit of robustness to tie robustness concepts
to a central ideal probability model---compare Definition~\ref{def:asy_breakdown_point}:
To get rid of
the dependence on possibly highly improbable sample configurations
leading to an overly small FSBP, but still
preserving the aspect of a finite sample, we propose an expected FSBP:
\begin{Def} \label{AFSBPDef}
For an estimator $T$ with FSBP $\ve_n^\ast=\ve_n^\ast(T; X_1,...,X_n)$,
we define the \textit{expected} FSBP or \textit{EFSBP} as
\begin{equation} \label{Expect}
\bar \ve_n^\ast(T) := \Ew \ve_n^\ast(T; X_1,...,X_n)
\end{equation}
where expectation is evaluated in the ideal model.
\end{Def}
At some places, if existent, for a sequence $T$ of estimators $T_n$, we
also consider the limit
\begin{equation}
\bar \ve^\ast(T) := \lim_{n\to \infty} \bar \ve_n^\ast(T_n)
\end{equation}
and which, for brevity, we also call EFSBP where unambigous.

Admittedly, the evaluation of the  expectation in \eqref{Expect} in general assumes
knowledge of the parameter, but some vague prior information could be used
to restrict the range of the plausible parameter values, say to $\xi \in (0.5;2)$,
and take the worst behavior of $\bar \ve_n^\ast(T)$ on this range to base our
decisions on, compare, e.g.\ Figure~\ref{ABPfig}.

\medskip

Weighted by their (ideal) occurrence probability, by this definition, improbable
sample configurations of the ideal sample---\textit{before} contamination---are
smoothed out in EFSBP; we still cannot exclude these
configurations, but usually by corresponding Che\-by\-shev-type inequalities
for growing sample size $n$ these will occur with decreasing probability
and $\ve_n^\ast$ will concentrate about $\bar \ve^\ast_n$.
Hence, in practice, without extra knowledge, \`a priori, the user can rely on being protected against
up to $\bar \ve_n^\ast(T) n$ outliers on average; i.e.; although there may be (rare) cases
where we have considerably less protection, these cases are
balanced by corresponding cases with considerably stronger protection.

By averaging, EFSBP is closer again to the ABP of
\citet{Ha:68}, but preserves the finite sample aspect of FSBP.
In the examples, we will show that this aspect is non-negligible,
and that for sample sizes about $40$, the ABP will still be
somewhat misleading (see Table~\ref{p0q1eTab} and Figure~\ref{EFSBPn} below), while at the same
time, as mentioned, FSBP will be way too pessimistic.
By dominated convergence though, the limit of EFSBP will coincide with the ABP
whenever the FSBP converges to the ABP.

\medskip
Small values of $\ve_n^\ast$ for particular samples do not only occur in
the models discussed here:
In the one-dimensional normal scale model, we can already have FSBP of $0$
for the median of absolute deviations \textit{MAD} for large enough values
of $n_0$ as introduced before \eqref{n0n1}. Such events
(and similarly extraneous sample configurations), however, occur
with probability $0$ in a continuous setting. Otherwise, in situations where
a FSBP of $0$ could occur with positive probability in the ideal model,
necessarily we have mass points violating the standard smoothness assumptions
usually required in scale models: the corresponding Fisher
information of scale would be infinite then, compare \cite{R:R:10}, and one
may then rather question the use of MAD. In our case, this is somewhat
different, as without arbitrary restrictions on the sample space,
samples with FSBP of $0$ can occur with small but positive ideal probability
(see $p_0$ in Table~\ref{p0q1eTab}), although our model remains smooth
(and Fisher information finite).

%
\section{Robust Estimators Types} \label{EstDef}

We illustrate the concept of FSBP in our scale-shape models
for Pickands-type and LD-type estimators, as defined in the sequel.

\subsection{Pickands Estimator} \label{PEsec}
%
\textit{Pickands estimator (PE)} for GPD is  a special case of the
Elementary Percentile Method (EPM) as discussed by \citet{C:H:97} for GPD.
Such estimators are based on the empirical quantiles, in our case, we follow
\citet{Pick:75} and use  the
empirical 50\%  and 75\%  quantiles $\hat Q_2$ and $\hat Q_3$. Pickands
estimators for $\xi$ and $\beta$ in GPD model then are defined as
\begin{align}
\hat{\xi} = \frac{1}{\log(2)}\log \frac{\hat Q_3 - \hat Q_2}{\hat Q_2}, \quad
\hat{\beta} = \hat{\xi}\,\frac{{\hat Q_2}^2}{\hat Q_3 - 2 \hat Q_2}
\end{align}
where we see that for $\hat\beta>0$ we have to require $\hat Q_3> 2\hat Q_2$,
in which case $\hat \xi>0$ automatically. Apparently  PE is equivariant in the sense of \eqref{scaleeq}.

\begin{figure}[!h]
\centering
\includegraphics[width=0.9\textwidth,height=0.4\textheight]{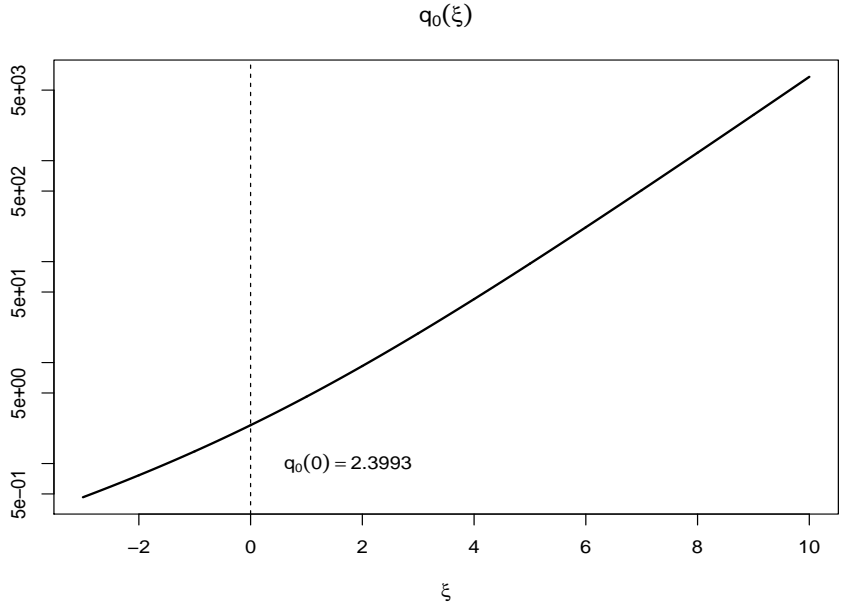}
\caption{\label{q0plot} $q_0(\xi)$ for different values of $\xi$; note the logarithmic $y$-scale
}
\end{figure}

For GEVD, analogue estimates can be obtained by
\begin{align}
&\hat{\xi} = \left \lbrace \xi \in \mathbb{R} \Large\mid \frac{\hat Q_3-\hat Q_2}{\hat Q_2} = q_0(\xi) \right \rbrace;\quad q_0(\xi)=\frac{\log(4/3)^{-\xi}-\log(2)^{-\xi}}{\log(2)^{-\xi}-1}, \\
&\hat{\beta} = \hat{\xi}\,\frac{{\hat Q_2}^2}{\hat Q_3 - 2 \hat Q_2} \frac{\log(4/3)^{-\hat\xi}+1-2\log(2)^{-\hat\xi}}{\log(2)^{-2\hat\xi}+1-2\log(2)^{-\hat\xi}}
\end{align}
where $q_0$ is obviously smooth, and, if plotted, easily seen to be strictly isotone, compare Figure~\ref{q0plot};
in particular, $\hat \xi>0$ iff $\hat Q_3 > \hat Q_2(1+q_0(0)) \doteq 3.39 \hat Q_2$, and
$\hat \beta >0$ iff $\hat Q_3 > 2 \hat Q_2$.

In the Weibull model, \cite{B:C:C:09} have shown Pickands (quantile) estimators to have an explicit representation as
\begin{align} \label{weibPick}
\hat{\xi} = \frac{f^{-1}_{1,1}(3/4) - f^{-1}_{1,1}(1/2)}{\log(\hat Q_3)-\log(\hat Q_2)},\quad
\hat{\beta} = \hat Q_2/(-\log(1/2))^{1/\hat{\xi}}
\end{align}
where $f^{-1}_{1,1}(\alpha) = \log(-\log(1-\alpha))$.

For the Gamma distribution the quantile estimates have no closed solutions, so the matching of
empirical and theoretical quantiles is to be done numerically by root solving procedures.
\subsection{MedkMAD and other LD estimators}\label{Sec:MedkMAD}
%
\textbf{L}ocation-\textbf{D}ispersion estimators, introduced by \cite{Ma:99}, match
empirical location and dispersion measures of data against their population counterparts
to get the estimates of model parameters, and are applicable for asymmetric location-scale (Lognormal),
as well as in scale-shape models (GPD, Pareto, Weibull, Gamma).

Let $\theta = (\alpha,\sigma)$ be a parameter vector, $F_n$, $F_{\alpha,\sigma}$ empirical
and model distribution functions, $m(F_n)$, $s(F_n)$, $m(F_{\alpha,\sigma})$,
$s(F_{\alpha,\sigma})$ corresponding empirical and model
location and dispersion, then LD estimators $(\hat{\alpha},\hat{\sigma})$ are solutions of \\
$$1) \quad \hat{\sigma} m(F_{0,1}) +\hat{\alpha}= m(F_n), \; \hat{\sigma} s(F_{0,1}) = s(F_n)$$
when $\alpha$ is a location parameter,
$$2) \quad \hat{\sigma} m(F_{\hat{\alpha},1}) = m(F_n), \; \hat{\sigma} s(F_{\hat{\alpha},1}) = s(F_n)$$
when $\alpha$ is a shape parameter.

Efficiency and robustness of these estimators depend on the choice of $m(\cdot)$ and $s(\cdot)$,
 and, of course, on the respective parametric model.
 Mean and standard deviation are classical measures for location and dispersion, respectively.
 Robust alternatives are median, trimmed mean---for location, IQR, MAD, trimmed MAD, Sn, Qn---for dispersion.
 In addition, for asymmetric distributions, we propose a new dispersion measure, namely kMAD.
 Table \ref{Tab:LD-estimators} displays different variations
 for LD estimators with increasing efficiency together with corresponding references.

\begin{table}[ht]
\begin{center}
\begin{tabular}{l|l|l}
Location & Dispersion & Location/Dispersion\\
\hline
\hline
Median
&\begin{tabular}{l} IQR \\
 (\textbf{I}nter\textbf{q}uantile \textbf{R}ange)
 \end{tabular}
&
\cite{Ma:99}
(Gamma, Weibull)
\\
\hline
 Median
& \begin{tabular}{l}MAD \\
(\textbf{M}edian of \textbf{A}bsolute \textbf{D}eviations)
 \end{tabular}
&
 \cite{B:C:C:09} (Weibull)\footnotemark \\
\hline
trimmed Mean &
\begin{tabular}{l}
trimmed M(ean)AD \\
\cite{Ma:99}
\end{tabular}
& \cite{Ma:99} (Gamma, Weibull)\\
\hline
Median
& \begin{tabular}{l}
kMAD\\
\cite{Ru:Ho:10}
\end{tabular}
& \cite{Ru:Ho:10} (GPD)\\
\hline
Median
&\begin{tabular}{l}
 $S_n$ \\
\cite{Ro:93}
\end{tabular}
& ---\\
\hline
Median
&\begin{tabular}{l}
 $Q_n$\\
\cite{Ro:93}
\end{tabular}
& \cite{B:C:C:09} (Weibull)\\
\end{tabular}
\caption{\label{Tab:LD-estimators}LD estimators and literature of using for scale-shape models}
\end{center}
\end{table}
\footnotetext{unchecked credit given to \cite{Oliv:06} in the cited reference}
\paragraph{Definitions of some particular LD estimators}
Empirical median $\hat m=\hat m_n$ and median of absolute deviations $\hat M=\hat M_n$ are well known for their high breakdown point, jointly achieving the highest possible asymptotic breakdown point of $50\%$ among all affine equivariant estimators at symmetric, continuous univariate distributions.

Hence it is plausible to define an estimator for $\xi$ and $\beta$, matching
$\hat m$ and $\hat M$ against their population counterparts $m$ and $M$
within a scale-shape model. It turns out that the mapping $(\beta,\xi)\mapsto (m,M)(F_\vartheta)$
is indeed a Diffeomorphism, hence for sufficiently large sample size $n$, we can solve
the implicit equations for $\beta$ and $\xi$ to obtain the MedMAD estimator.

More efficient estimators for dispersion than MAD, but with same
breakdown point of $50\%$ at continuous distributions, and in particular
suitable for asymmetric distributions, have been proposed in \cite{Ro:93} as $\hat M = Q_n$ and
$\hat M = S_n$. In this context,
$Q_n = \{|x_i-x_j|; \ i<j \}_{(k)}$, $k = \binom{h}{2} \approx \binom{n}{2}/4$, $h = \lfloor n/2\rfloor + 1$,
while $S_n = \textrm{med}_i \{ \textrm{med}_j |x_i-x_j| \}$ where in case of discrepancies, the inner median
is to be taken as \textrm{hi-med}, the outer as \textrm{lo-med}, where
$\textrm{lo-med}(F)=F^{-}(1/2)$, and $\textrm{hi-med}(F)=F^-(1/2+{\scriptstyle 0})$.
The resulting LD estimators are named MedQn and MedSn,
respectively.

Note that for asymmetric  $G$, the functionals
$S(G)= \rm med_X \ med_Y|X-Y|, \ X,Y \sim G$ and $Q(G)= \inf\{s>0; \int G(t+d^{-1}s)dG(t)\geq 5/8\}$
involve expensive, careful numerical calculations, in particular for the heavy-tailed GPD and GEVD cases.

In the GEVD and GPD case, due to their considerable skewness to the right, one can improve the
MedMAD estimator considerably, using a dispersion functional that takes this skewness
into account: For a distribution $F$ on $\R$ with median $m$
let us define for $k>0$
\begin{equation}
{\rm kMAD}(F,k):=\inf\big\{\,t>0\,\big|\, F(m+kt)-F(m-t)\ge 1/2\,\big\}
\end{equation}

i.e.; kMAD only searches among the class of intervals about the median
 $m$ with covering probability $50\%$, where the part right to $m$ is $k$
 times longer than the one left to $m$ and returns the shortest of these.
In our case, $k$ would be chosen
to be a suitable number larger than $1$, and $k=1$ would reproduce the MAD.
 Apparently, whenever $F$ is continuous, kMAD preserves the ABP of the
MAD of $50\%$, i.e.; covering both the explosion and implosion case.


\paragraph{Computation of LD estimators} Each of our
dispersion estimators Sn, Qn, and kMAD is scale-equivariant, and
the same also holds for the respective population counterparts,
as well as for any fixed quantile,
in particular for the median; hence denoting the dispersion functional
by $s$, both the quotient
   $q(\xi):=s(\beta,\xi)/m(\beta,\xi,)$
and its empirical counterpart $\hat q_{n}$ ($q_k,\hat q_{k;n}$ for MedkMAD)
are scale-free; so we have reduced the problem by one dimension.
In the sequel we also write $q_k$, $\hat q_{k;n}$ for Sn and Qn,
where $k$ is then simply void.
Assuming continuity and monotonicity, we obtain an estimator for
$\xi$ given by $\hat \xi_n=q^{-1}_k(\hat q_{n,k})$.

A corresponding estimator for $\beta$ for each of the variants
kMAD, Sn, and Qn, is then simply given by
\begin{equation}
\hat \beta_n=\hat m / m(1,\hat \xi_n)
\end{equation}

In particular, by construction all LD estimators are equivariant in the sense
of \eqref{scaleeq}.

\paragraph{Continuity and Monotonicity}
of $q$ as a function in $\xi$ ensure existence and uniqueness of
the implicitly defined estimator for $\xi$.

Continuity of  $q_k$  in $\xi$ for
all our scale-shape models, i.e.; GPD, GEVD, Gamma, and Weibull
and all our dispersion functionals ${\rm kMAD}(k)$, $S$ and $Q$
is straightforward, even for the limit cases $\xi\to0$.

Monotonicity of $q_k$, though, is not so obvious from the analytic terms,
but the plots 
of function $\xi\mapsto q(\xi)$  for dispersions kMAD, Sn, and Qn,
in Figure~\ref{fig:LD_quotients} indicate strict monotonicity for each
of the
dispersions and the GPD, Gamma, and Weibull cases, while for
the GEVD case, $q$ is bitone with maximum $\bar q_k$ taken in $\xi_0>0$.
To obtain consistent estimators in this case, we restrict ourselves
to the range left or right to $\xi_0$ containing $\xi=0.7$ in this paper.

\paragraph{Restriction(s) of solvability domain}
Besides this restriction of the range of $\xi$ in the GEVD case,
we conclude, that in
the GPD and in GEVD cases, for each of the dispersions, our restriction to  $\xi>0$ implies
a restriction of the solvability domain for $q_k(\xi)$ with
in the set of admissible values of $\xi$:
\begin{equation}
q_k(\xi)\ge \lim_{\xi\to 0} q_k(\xi)=:\check q_k>0
\end{equation}
while in the Weibull and Gamma case, $\check q_k$ can be taken
as $0$.

The following lemma gives us yet other restrictions:
\begin{Lem}\label{Lem2}
Let $s$ the functional version to any of the scale estimators Sn, Qn,
and ${\rm kMAD}$ (for any $k>0$).
Let $G$ be a distribution  on $\R$ such that $-\infty<x_0=\sup\{x\colon\;G(x)=0\}$, i.e.; with finite left endpoint. Then with $m=G^{-}(1/2+{\scriptstyle 0})$,
the hi-med of $G$,
\begin{equation}
s(G)\leq m-x_0=:s_0
\end{equation}
with equality iff
\begin{description}
\item[\parbox{1cm}{(kMAD)}] $\quad$ $G((m;m+ks_0))=0$.
\item[\parbox{1cm}{(Sn)}] $\quad$ $G(x+2s_0 - {\scriptstyle 0})-G(x) <1/2$ for each $x\ge x_0$.
\item[\parbox{1cm}{(Qn)}] $\quad$ $G(m)=1/2$, $G(x_0)=0$.
\end{description}

\end{Lem}

Consequently, as $x_0=0$, in the GPD, Gamma, and Weibull case,
\begin{equation}
q_k(\xi)< 1\qquad \forall \xi
\end{equation}
and, the same relation in the ideal model also holds sample-wise, i.e.;
\begin{equation}
\hat q_{k,n}< 1=:\bar q_k
\end{equation}
in each sample (from the ideal model distribution) where
\begin{description}
\item[\parbox{1.3cm}{(kMAD)}] at least one observation in $\big(\hat m; \hat m + k (\hat m- X_{(1)}) \big)$.
\item[\parbox{1.3cm}{(Sn)}] at least one interval of length shorter than $2(\hat m - X_{(1)})$ containing more
\ifx\blinded\undefined
\hphantom{$\mbox{\hspace{.9cm}}$}
\fi than  $\lfloor n/2\rfloor +1 $ observations.
\item[\parbox{1.3cm}{(Qn)}] all observations finite.
\end{description}

Hence, for the LD estimators, we have to find
the unique zero $\hat \xi_n$ of $H_k(\xi)=q_k(\xi)-\hat q_{n,k}$ in the
interval $(\check q_k;\bar q_k)$ which can easily be solved  with a standard
univariate root-finding tool like {\tt uniroot} in {\sf R} \citep{R}.

\paragraph{Producing breakdown}
Clearly, in the GPD case, we could drive $\hat q_{k,n}$ to values larger than $1$
by modifying observations in the original sample to values smaller
than $x_0$. These values would then be identifiable as outliers
without error then, and we could cancel them from the sample.
Instead we only consider contaminations by values larger than $x_0$
(which could also have been produced in the ideal model).

On first glance, values of $\hat q_{k,n}$ outside $(\check q_k, \bar q_k)$
would make for a ``definition breakdown'', but if, for
 $\hat s_n$ the respective scale estimator, $\hat s_n \to \hat m$,
 this entails $\hat \xi_n \to \infty$  in the GPD case
and  $\hat \xi_n \to 0$ in the Gamma and Weibull case. Hence
 we can produce a breakdown in the original sense
by modifying an original sample such that $\hat s_n \to \hat m$.


\section{Calculation of (E)FSBP for Pickands and LD Estimators} \label{EstBP}
In some of our scale-shape models and for some of our estimators
we have analytic expressions for the different breakdown point notions.
\subsection{Pickands Estimator} \label{PEsecBP}
\begin{Prop}[Breakdown for PE] \label{PEprop}
In the GPD, GEVD, Weibull, and Gamma cases, an upper bound for FSBP of PE is given by
$25\%$, which also invariably is the FSBP in the Weibull case.
In the GPD case, no matter if $\xi\in\R$ or $\xi>0$, and in the unrestricted GEVD case, i.e.; $\xi\in\R$, FSBP is given by
\begin{equation} \label{Nn0def}
\ve_n^\ast = \hat N^0_n /n,\qquad\mbox{for } \quad\hat N^0_n:=\#\{X_i\,\big|\, 2\hat Q_2\leq X_i\leq \hat Q_3\}.
\end{equation}
The ABP then is given by
\begin{equation}
\bar \ve^\ast=\ve^\ast= P_{\vartheta}(2Q_2<X_1\leq Q_3)
\end{equation}
which in the GPD case is just $\bar \ve^\ast=(2^{\xi+1}-1)^{-1/\xi}-1/4$,
and, in the GEVD case, $\bar \ve^\ast = 3/4-\exp\Big(-\big(2 \log(2)^\xi-1\big )^{-1/\xi}\Big)$.
In the restricted GEVD case, where $\xi>0$,
\begin{equation} \label{Nn0def2}
\ve_n^\ast = \tilde N^0_n /n,\qquad\mbox{for } \quad\tilde N^0_n:=\#\{X_i\,\big|\, q_0(0) \hat Q_2\leq X_i\leq \hat Q_3\}.
\end{equation}
The ABP then is given by
\begin{equation}
\bar \ve^\ast=\ve^\ast= P_{\vartheta}(q_0(0) Q_2<X_1\leq Q_3).
\end{equation}
\end{Prop}
For $\xi=0.7$,  we obtain
$\bar \ve^\ast\doteq 6.42\%$ in the GPD case, and in the GEVD case, $\bar \ve^\ast\doteq 15.42\%$
in the unrestricted case, and $\bar \ve^\ast\doteq 6.13\%$ in the restricted case. %
For the figures for $\bar \ve^\ast_n$, for $n=40,100,1000$ in the GPD, GEVD,
and Weibull case, see Table~\ref{Tab:n40M10000_neg}, where we make use of
Proposition~\ref{Nnprop} below.
In the  Gamma case, the situation is more involved, and we skip computation of the actual breakdown points.

%

\subsection{LD Estimators} \label{MedkMADsecBP}

The FSBPs of $50\%$ of the median and the dispersion estimators
obviously form an upper bound for the FSBP of
the LD estimators, implying that you could at least drive one of the parameters
$\beta$ and $\xi$ to $\infty$. However, similarly to regression based estimators for the Weibull case
of \citet{B:C:C:09}, breakdown is not only entailed by moving mass
to $0$ or $\infty$, and the actual breakdown points
of the LD estimators are smaller; for the MedkMAD, we come up with some
explicit expressions, while for the MedSn and MedQn we have to recur to
simulations, see Subsection~\ref{simusec}.



\begin{figure}[!h]
\centering
\includegraphics[width=0.9\textwidth,height=0.4\textheight]{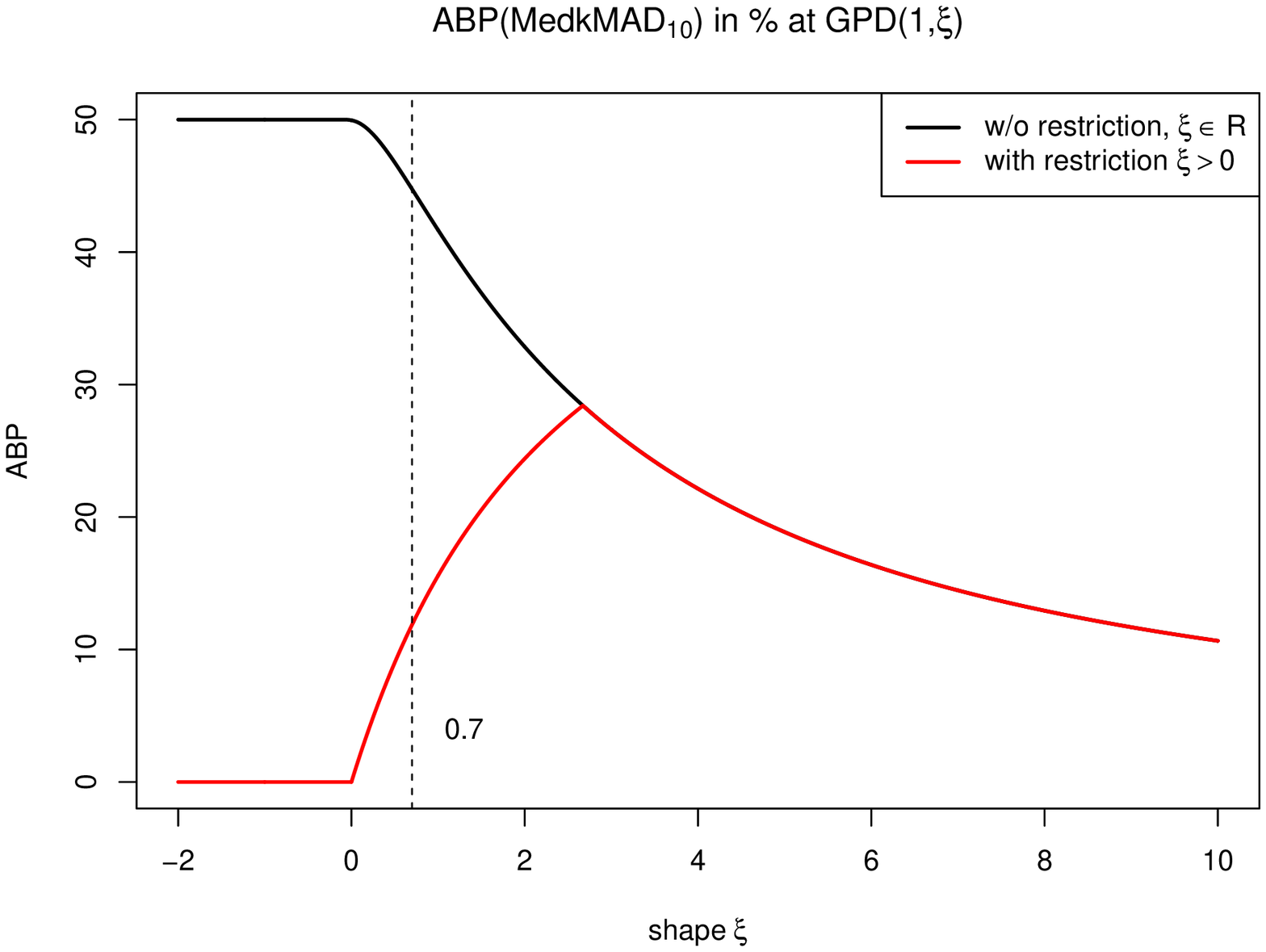}
\caption{\label{ABPfig} $\bar \ve^\ast({\rm MedkMAD}_{10};{\rm GPD}_{\vartheta=(1,\xi)})$ for different $\xi$ with or without
restriction $\xi>0$
}
\end{figure}

\begin{Prop}[Breakdown for MedkMAD] \label{MedkMADprop}
In the GPD, Weibull, and Gamma cases, the FSBP of MedkMAD is given by
\begin{eqnarray}
\ve_n^\ast &=& \left\{\begin{array}{ll}
\hat N_n'/n & \quad\mbox{Weibull; Gamma; GPD, unrestr. case, i.e.; } \xi \in\R\\
\min(\hat N_n',\hat N_n'')/n& \quad\mbox{GPD, restr. case, i.e.; } \xi>0
\end{array} \right.\quad\\
\hat N_n'&:=&\#\{X_i \,| \hat m < X_i\leq (k+1)\hat m \,\},\label{Nn1def}\\
\hat N_n''&:=&\lceil n/2\rceil - \#\{X_i \,|\, (1-\check q_k) \hat m < X_i
< (k\check q_k+1)\hat m \}\label{Nn2def}.
\end{eqnarray}
The ABP in this case is given by $\bar \ve^\ast= \bar \ve'$ for the unrestricted
and $\bar \ve^\ast= \min (\bar \ve', \bar\ve'')$ for the restricted case  where
\begin{equation}
\bar \ve' = F_\vartheta((k+1)m)- 1/2, \qquad
\bar \ve'' =1/2-F_\vartheta\big((k\check q_k+1)m\big)+ F_\vartheta\big((1-\check q_k)m\big).
\end{equation}
\end{Prop}
At $k=10$ and $\xi=0.7$, we obtain
$\bar \ve^\ast\doteq 44.75\%$ (GPD; $\xi\in\R$),  $11.87\%$ (GPD; $\xi>0$), $49.47\%$ (Gamma), and $47.56\%$ (Weibull). %
For further figures for $\ve^\ast_n,\bar \ve^\ast_n,\bar \ve$, see Table~\ref{Tab:n40M10000_neg},
where again we make use of Proposition~\ref{Nnprop}.
In particular, contrary to \cite{B:C:C:09}, not only is our
FSBP varying sample-wise in these cases, but also do ABP and EFSBP depend on
$\xi$.
A plot of the dependency $\xi\mapsto \bar \ve^\ast({\rm MedkMAD}_{10};{\rm GPD}(\xi))$
is displayed in Figure~\ref{ABPfig}.

\subsection{Calculation of EFSBP}
\begin{figure}[!h]
\centering
\includegraphics[width=0.9\textwidth,height=0.4\textheight]{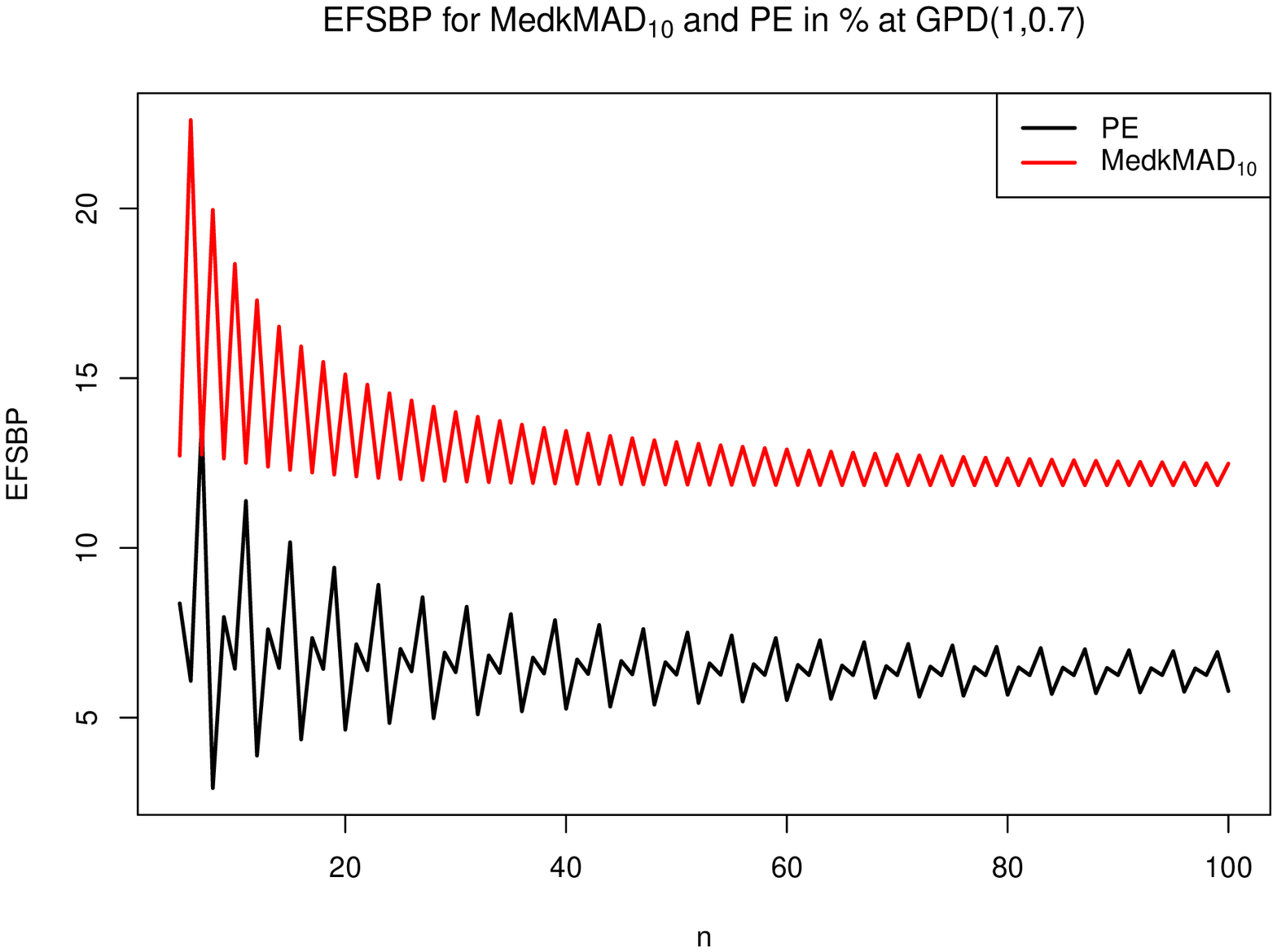}
\caption{\label{EFSBPn} $\bar \ve^\ast_n$ for PE and ${\rm MedkMAD}_{10}$ at ${\rm GPD}_{(1,0.7)}$ (restricted to $\xi>0$) as a function in $n$
}
\end{figure}

To obtain actual values of EFSBP, we have the following proposition.
\begin{Prop}\label{Nnprop}
Consider $\hat N^0_n$, $\hat N'_n$, $\hat N''_n$ as defined in
\eqref{Nn0def}, \eqref{Nn1def}, \eqref{Nn2def}
and write $\bar F$ for $1-F$. Then for $n\ge 3$,
\begin{ABCTH}
\item $\!\!\!\!\!\!\!\!\!$ setting  $i_1=\lfloor n/2\rfloor$, $i_2=\lceil 3n/4\rceil$, and abbreviating
$2F^{-1}(u)$ by $t_2$, we obtain for $l\in\{1,\ldots,i_2-i_1-1\}$
\begin{equation}
P(\hat N^0_n= l)=
n \int_{0}^1 \!\!\! {\textstyle {n-1 \choose i_1-1,i_2-i_1-l-1}} u^{i_1-1}
   \big(F(t_2)-u\big)^{i_2-i_1-l-1} \bar F(t_2)^{n-i_2+l+1}\,du \label{NhatPE}
\end{equation}
and
\begin{equation}
P(\hat N^0_n= 0)=
n \sum_{l=0}^{n-i_2} \int_{0}^1 \!\!\! {\textstyle {n-1 \choose i_1-1,i_2-i_1+l}} u^{i_1-1}
   \big(F(t_2)-u\big)^{i_2-i_1+l} \bar F(t_2)^{n-i_2-l}\,du \label{NhatPE0}.
\end{equation}
The case of $\tilde N^0_n$ is obtained from \eqref{NhatPE}, \eqref{NhatPE0} replacing $t_2$ by $t_q:=q_0(0)F^{-1}(u)$.
\item $\!\!\!\!\!\!\!\!\!$ using the hi-med and setting $t_k:=(k+1)F^{-1}(u)$, we obtain
for $l\in\{0,\ldots,\lceil n/2\rceil-2\}$
\begin{equation}
P(\hat N_n'= l) =
n \int_{0}^1 \!\!\!{\textstyle {n-1 \choose \lfloor n/2\rfloor+1 ,l}} u^{n/2}
 \big(F(t_k)-u\big)^{l} \bar F(t_k)^{n/2-1-l}\,du \label{NhatMM}
\end{equation}
\item $\!\!\!\!\!\!\!\!\!$ setting $t_+:=(1+k\check q_k)F^{-1}(u)$, $t_-:=(1-\check q_k)F^{-1}(u)$, we obtain
for $l\in\{0,\ldots,n/2-1\}$
\begin{eqnarray}
P(\hat N_n''= n/2-l)&=&
n \sum_{l_2=0}^l {\textstyle {n-1 \choose n/2-l_2-1,l_2,l-l_2}}
\int_{0}^1 F(t_-)^{n/2-l_2-1}
\big(u-F(t_-)\big)^{l_2} \times\nonumber\\
&&\qquad\times \big(F(t_+)-u\big)^{l-l_2} \big(1-F(t_+)\big)^{n/2+l_2-l}
\,du \label{NhatkMeM2}.
\end{eqnarray}
\end{ABCTH}
\end{Prop}
The dependency of EFSBP on $n$ is visualized in Figure~\ref{EFSBPn}.
We see a saw-tooth like oscillation which is explained by the
use of finite sample quantiles in Proposition~\ref{Nnprop}. In particular
there are considerable deviations from ABP for moderate sample sizes.


\subsection{Illustration: Usefulness of EFSBP}
The expressions given in Propositions~\ref{PEprop},
\ref{MedkMADprop}, and \ref{Nnprop} illustrate that in both the Pickands and LD
estimator case, even starting from an ideal sample, the
``usual'' sample-wise fluctations of ${\rm FSBP}=\hat N_n/n$
are considerable. Moreover, Proposition~\ref{Nnprop}
shows that we even have a positive,
although very small ideal probability
\begin{equation}
p_0:=P^X(\hat N_n=0)>0
\end{equation} for breakdown already in the ideal model.
Now, on the event $\{\hat N_n=0\}$, $\ve_n^\ast=0$, so no universal non-trivial lower bound can be given
for the FSBP in both the Pickands and LD estimator case.
As the figures in Table~\ref{p0q1eTab} below illustrate, however,
such an event will hardly ever occur provided only moderately small sample
sizes, and the same goes for similarly small realizations
of $\hat N_n$, so these cases, as motivated in the introduction of EFSBP,
are not representative, indeed.
To grasp the difference between $\bar \ve_n^\ast$ and $\ve^\ast$, we consider the following
Hoeffding-type lemma for empirical quantiles
\begin{Lem} \label{HoeffLemm}
\begin{itemize}
\item[(a)]
Let $0<\delta<1/2$ and $t \in \R$ and for given $\alpha\in(0,1)$ and cdf $F$, let $q=F^{-}(\alpha)$,
and $\hat q_n=\hat F_n^{-}(\alpha)$. Assume that $F$ is differentiable in $q$ with density
$f(q)>0$. Then with $t_n=t n^{-1/2+\delta}$, for $n$ large enough,
\begin{equation}
P(|\hat q_n-q|\ge t_n) \leq \exp(-2 f(q)^2 n^{\delta})
\end{equation}
\item[(b)] Let $a_i\not=0$, $\alpha_i\in(0,1)$, $\alpha_1\not=\alpha_2$ $i=1,2$ be given as well as
cdf $F$; assume $F$ differentiable in  $a_iq_i$, $i=1,2$. Then under the assumptions of (a) for $q_i$,
for $\hat I_n=(a_1 \hat q_{1,n},a_2 \hat q_{2,n})$
and $I=(a_1q_1; a_2q_2)$, we have for $n$ large enough,
\begin{equation}
P^X(\hat I_n) = P^X(I) + \LO(n^{-1/2+\delta/2}).
\end{equation}
\end{itemize}
\end{Lem}
To illustrate the size of the $\LO(n^{-1/2+\delta/2})$-term,
let us also determine the upper $p_1$-quantile
of $\ve_n^\ast$ for $p_1=0.95^{0.0001}$, i.e.; the minimal number $q_1$,
such that with probability $0.95$ we will not see realizations with
$\ve_n^\ast<q_{1}$ in $10000$ runs of sample size $n$.

\paragraph{Evaluations for PE and MedkMAD} Using the actual distribution of $\hat N_n$ given in Proposition~\ref{Nnprop},
in Table~\ref{p0q1eTab}, for  Pickands (PE) and MedkMAD, $k=10$ we determine $\bar \ve_n^\ast$, $p_0$ and $q_1$
for $n=40, 100, 1000$ in the GPD (with and without restriction to $\xi>0$), Gamma, and Weibull cases, each with $\xi=0.7$.
The Gamma case is skipped, though, in the PE case for lack of explicit formulae. 
Apparently $\bar \ve_n^\ast$ is quickly converging in $n$, so $\bar\ve^\ast$
gives indeed a useful bound on average.

\begin{table}[!t]
\begin{center}\small
GPD\\[1ex]
\begin{tabular}{l|c|r@{\hspace{.2mm}}l@{\hspace{.1mm}}rr@{\hspace{.2mm}}l@{\hspace{.1mm}}rr@{\hspace{.2mm}}l@{\hspace{.1mm}}rr@{\hspace{.2mm}}l@{\hspace{.1mm}}rr}
\hline
                &estimator      & \multicolumn{3}{c}{$n = 10$} & \multicolumn{3}{c}{$n = 40$} & \multicolumn{3}{c}{$n = 100$}   & \multicolumn{3}{c}{$n = 1000$} &$n=\infty$        \\
\hline
$p_0$           &PE            & $5.1$ & ${\rm e}-$ & $01 $  & $2.7$ & ${\rm e}-$ & $01 $ & $ 7.9$ & ${\rm e}-$ & $02 $ & $ 5.4$&${\rm e}-$ & $08$&0\\
                &MedkMAD, $\xi\in\R$      & $3.3$ & ${\rm e}-$ & $04 $  & $1.6$ & ${\rm e}-$ & $15 $ & $ 7.2$ & ${\rm e}-$ & $38 $ & $ <1 $&${\rm e}-$ & $300$&0\\
                &MedkMAD, $\xi>0$       & $1.4$ & ${\rm e}-$ & $01 $  & $3.5$ & ${\rm e}-$ & $02 $ & $ 2.7$ & ${\rm e}-$ & $03 $ & $ 2.9$&${\rm e}-$ & $018$&0\\
\hline
$q_1$           &PE            & \multicolumn{3}{r}{$ 0.00\%$}  & \multicolumn{3}{r}{$ 0.00\%$} & \multicolumn{3}{r}{$ 0.00\% $} & \multicolumn{3}{r}{$ 1.00\%$}& $ 6.42\%$\\
                &MedkMAD, $\xi\in\R$      & \multicolumn{3}{r}{$ 0.00\%$} & \multicolumn{3}{r}{$ 20.00\%$} & \multicolumn{3}{r}{$30.00\% $} & \multicolumn{3}{r}{$41.10\%$}& $44.75\%$\\
                &MedkMAD, $\xi>0$       & \multicolumn{3}{r}{$ 0.00\%$} &\multicolumn{3}{r}{$ 0.00\%$} & \multicolumn{3}{r}{$0.00\% $} & \multicolumn{3}{r}{$5.70\%$}& $11.87\%$\\
\hline
$\bar\ve^\ast_n$&PE            & \multicolumn{3}{r}{$ 6.44\%$}& \multicolumn{3}{r}{$ 5.26\%$} & \multicolumn{3}{r}{$ 5.78\% $} & \multicolumn{3}{r}{$ 6.34\%$}& $ 6.42\%$\\
                &MedkMAD, $\xi\in\R$     & \multicolumn{3}{r}{$35.85\%$}  & \multicolumn{3}{r}{$42.53\%$} & \multicolumn{3}{r}{$43.86\% $} & \multicolumn{3}{r}{$44.66\%$}& $44.75\%$\\
                &MedkMAD, $\xi>0$      & \multicolumn{3}{r}{$18.37\%$}  & \multicolumn{3}{r}{$13.45\%$} & \multicolumn{3}{r}{$12.48\% $} & \multicolumn{3}{r}{$11.94\%$}& $11.87\%$\\
\hline
\end{tabular}\\[1.5ex]
GEVD\\[1ex]
\begin{tabular}{l|c|r@{\hspace{.2mm}}l@{\hspace{.1mm}}rr@{\hspace{.2mm}}l@{\hspace{.1mm}}rr@{\hspace{.2mm}}l@{\hspace{.1mm}}rr@{\hspace{.2mm}}l@{\hspace{.1mm}}rr}
\hline
                &estimator      & \multicolumn{3}{c}{$n = 10$} & \multicolumn{3}{c}{$n = 40$} & \multicolumn{3}{c}{$n = 100$}   & \multicolumn{3}{c}{$n = 1000$} &$n=\infty$        \\
\hline
           &\hphantom{MedkMAD, $\xi\in\R$}& \multicolumn{3}{r}{}& \multicolumn{3}{r}{}& \multicolumn{3}{r}{}& \multicolumn{3}{r}{}            &  \\[-2ex]
$p_0$           &PE, $\xi\in\R$            & $2.8$ & ${\rm e}-$ & $01 $  & $3.8$ & ${\rm e}-$ & $02 $ & $ 6.8$ & ${\rm e}-$ & $04 $ & $ 8.2$&${\rm e}-$ & $28$&0\\
                &PE, $\xi>0$            & $5.4$ & ${\rm e}-$ & $01 $  & $3.7$ & ${\rm e}-$ & $01 $ & $ 2.0$ & ${\rm e}-$ & $01 $ & $ 5.0$&${\rm e}-$ & $04$&0\\
\hline
$q_1$           &PE, $\xi\in\R$    & \multicolumn{3}{r}{$ 0.00\%$}  & \multicolumn{3}{r}{$ 0.00\%$} & \multicolumn{3}{r}{$ 0.00\% $} & \multicolumn{3}{r}{$ 9.10\%$}& $ 15.42\%$\\
                &PE, $\xi>0$            & \multicolumn{3}{r}{$ 0.00\%$}  & \multicolumn{3}{r}{$ 0.00\%$} & \multicolumn{3}{r}{$ 0.00\% $} & \multicolumn{3}{r}{$ 0.00\%$}& $ 6.13\%$\\
\hline
$\bar\ve^\ast_n$&PE, $\xi\in\R$            & \multicolumn{3}{r}{$ 12.50\%$}& \multicolumn{3}{r}{$ 14.38\%$} & \multicolumn{3}{r}{$ 14.78\% $} & \multicolumn{3}{r}{$ 15.33\%$}& $ 15.42\%$\\
                &PE, $\xi>0$            & \multicolumn{3}{r}{$ 4.80\%$}& \multicolumn{3}{r}{$ 5.54\%$} & \multicolumn{3}{r}{$ 6.04\% $} & \multicolumn{3}{r}{$ 6.09\%$}& $ 6.13\%$\\
\hline
\end{tabular}\\[1.5ex]
Gamma\\[1ex]
\begin{tabular}{l|c|r@{\hspace{.2mm}}l@{\hspace{.1mm}}rr@{\hspace{.2mm}}l@{\hspace{.1mm}}rr@{\hspace{.2mm}}l@{\hspace{.1mm}}rr@{\hspace{.2mm}}l@{\hspace{.1mm}}rr}
\hline
                &estimator      & \multicolumn{3}{c}{$n = 10$} & \multicolumn{3}{c}{$n = 40$} & \multicolumn{3}{c}{$n = 100$}   & \multicolumn{3}{c}{$n = 1000$} &$n=\infty$        \\
\hline
$p_0$           &MedkMAD       & $2.3$ & ${\rm e}-$ & $04 $ & $2.7$ & ${\rm e}-$ & $14 $ & $ 4.8$ & ${\rm e}-$ & $34 $ & $ <1 $&${\rm e}-$ & $300$&0\\
\hline
$q_1$           &MedkMAD       & \multicolumn{3}{r}{$ 0.00\%$} & \multicolumn{3}{r}{$ 22.50\%$} & \multicolumn{3}{r}{$38.00\% $} & \multicolumn{3}{r}{$47.60\%$}& $49.47\%$\\
\hline
$\bar\ve^\ast_n$&MedkMAD     &  \multicolumn{3}{r}{$39.03\%$} & \multicolumn{3}{r}{$46.80\%$} & \multicolumn{3}{r}{$48.40\% $} & \multicolumn{3}{r}{$49.37\%$}& $49.47\%$\\
\hline
\end{tabular}\\[1.5ex]
Weibull\\[1ex]
\begin{tabular}{l|c|r@{\hspace{.2mm}}l@{\hspace{.1mm}}rr@{\hspace{.2mm}}l@{\hspace{.1mm}}rr@{\hspace{.2mm}}l@{\hspace{.1mm}}rr@{\hspace{.2mm}}l@{\hspace{.1mm}}rr}
\hline
                &estimator      & \multicolumn{3}{c}{$n = 10$} & \multicolumn{3}{c}{$n = 40$}& \multicolumn{3}{c}{$n = 100$}   & \multicolumn{3}{c}{$n = 1000$} &$n=\infty$        \\
\hline
$p_0$           &PE           & \multicolumn{3}{r}{0} & \multicolumn{3}{r}{0}& \multicolumn{3}{r}{0}& \multicolumn{3}{r}{0}&0\\
                &MedkMAD      & $6.4$ &${\rm e}-$ & $04 $  & $5.5$ & ${\rm e}-$ & $13 $ & $ 5.6$ & ${\rm e}-$ & $31 $ & $ <1 $&${\rm e}-$ & $300$&0\\
\hline
$q_1$           &PE            & \multicolumn{3}{r}{$ 25.00\%$} & \multicolumn{3}{r}{$ 25.00\%$} & \multicolumn{3}{r}{$ 25.00\% $} & \multicolumn{3}{r}{$ 25.00\%$}& $ 25.00\%$\\
                &MedkMAD       & \multicolumn{3}{r}{$ 0.00\%$} & \multicolumn{3}{r}{$ 17.50\%$} & \multicolumn{3}{r}{$32.00\% $} & \multicolumn{3}{r}{$44.20\%$}& $47.56\%$\\
\hline
$\bar\ve^\ast_n$&PE            & \multicolumn{3}{r}{$ 25.00\%$} & \multicolumn{3}{r}{$ 25.00\%$} & \multicolumn{3}{r}{$ 25.00\% $} & \multicolumn{3}{r}{$ 25.00\%$}& $ 25.00\%$\\
                &MedkMAD       & \multicolumn{3}{r}{$37.68\%$}  & \multicolumn{3}{r}{$45.03\%$} & \multicolumn{3}{r}{$46.54\% $} & \multicolumn{3}{r}{$47.46\%$}& $47.56\%$\\
\hline
\end{tabular}
\caption{\label{p0q1eTab} {$p_0$, $q_1$, and $\bar\ve^\ast_n$ for PE and
MedkMAD ($k=10$)}}
\end{center}
\end{table}

According to the values of $p_0$, breakdown in the ideal model will hardly ever happen for PE
for $n\ge 1000$, and for MedkMAD for $n\ge 100$, and only rarely for $n\ge 40$.

The values for $q_1$ demonstrate that in a simulation study at the GPD with $\xi=0.7$
with $10000$ runs of sample size upto  $n=1000$, we will
probably see breakdowns for PE, as well as
for the MedkMAD restricted to $\xi>0$.
Contrary to this, as long as we have no more outliers than $8$, $30$, $411$ for sample sizes
$n=40,100,1000$, we will not see a breakdown for MedkMAD in the unrestricted case;
in the Gamma case with same shape we obtain $9$, $38$, $476$,
and in the Weibull $7$, $32$, $442$; analogue figures for PE at the Weibull with $\xi=0.7$
are $10$, $25$, $250$.

We may interpret the values of $\bar \ve_n$ as follows: Before having made any
observations, at the GPD at $\xi=0.7$, using PE, one may be confident
to be protected against $3$ outliers for sample size $40$, $7$ for
sample size $100$, and $65$ for sample size $1000$, while for MedkMAD,
the corresponding figures are $17$, $43$, and $447$ in the unrestricted case
and $5$, $12$, and $118$ when restricted to $\xi>0$; calculations
in the Gamma and Weibull cases give comparable numbers.
%
\subsection{Breakdown Calculations in the Remaining Cases: Simulational Approach}\label{simusec}

For the breakdown point of MedQn and MedSn, as well as for MedkMAD in the GEVD case,
there are no analytical expressions, so we calculate them using simulations.

More precisely, for each of the estimators MedkMAD ($k=10$), MedQn, MedSn, PE,
and each of the ideal distributional settings
GPD, GEVD, Weibull, and Gamma (each at $\vartheta=(1,0.7)$), we produced
$M=10000$ runs of sample sizes $n=40,100,1000$ and noted the number of
alterations needed to move $\hat q_{k,n}$ to $\bar q$, and in a second round,
starting from the same runs of ideal observations,
for GPD and GEVD, the minimal number of alterations needed to move $\hat q_{k,n}$ to $\check q_k$,
respectively the minimum of these two rounds. In the cases where explicit
formulae are available this gives us a possibility to cross-check our results.
Some small discrepancies should arise though, as we use the default median in {\sf R}, \cite{R},
i.e.; $(\textrm{hi-med}+\textrm{lo-med})/2$ for even sample size, while Proposition~\ref{Nnprop}
below is limited to $\textrm{hi-med}$.
For actual simulated values for $\bar \ve^\ast_n$, see
Table~\ref{Tab:n40M10000_neg}.

\section*{Conclusion}
This article provides a new measure for global robustness of an estimator at finite samples, i.e.;
EFSBP, a variant of the finite sample breakdown point which is particularly useful in situations where
we have only partial equivariance and no non-trivial, universal lower bounds for FSBP are available. This variant
comes closer to the (sample-free) ABP while still retaining the finite sample aspect of FSBP.

 We have illustrated this measure at a set of scale-shape models, applying it to LD and Pickands/Quantile-type estimators
 meant for high-breakdown initial estimators to be enhanced in efficiency by reweighting afterwards.

Although kMAD, Qn, and Sn all share the same
breakdown properties in the location-scale setting, where they are defined,
the corresponding LD estimators in the considered scale-shape models exhibit a differentiated
breakdown behavior, and  there is not one single best estimator.

In the unrestricted GEVD case, the easy-to-compute Pickands-type estimator
turned out to have the highest breakdown point
among all considered estimators, while in the setting restricted to $\xi>0$,
from sample size $100$, MedkMAD becomes superior. In all other situations, the best
estimator is either MedkMAD or MedQn. In the unrestricted and restricted
GPD case MedQn performs best, with MedkMAD close in the unrestricted case for $n=40$.
In the Weibull and Gamma cases MedkMAD performs best, except for the Weibull at $n=1000$ where
MedQn is best, but with MedkMAD close by. For deciding between MedkMAD and MedQn in cases
where their breakdown points are similar though, one also should take into
account computational costs as well, which so far clearly favors MedkMAD.

\appendix
\section{Proofs}\label{proofsec}

\begin{proof} { to Lemma~\ref{Lem2}: }
For any $k>0$, $G(m+ks_0)-G(m-s_0-{\scriptstyle 0})=G(m+ks_0)\ge 1/2$, so $s_0\geq {\rm kMAD}(G,k)$.
For $x\ge x_0$ and $Y\sim G$, let $g_G(x)={\rm med}_x(|Y-x|)=\inf\{s\ge 0\colon\;G(s+x)-G(x-s-{\scriptstyle 0})\ge 1/2\}$.
But $G(s_0+x)-G(x-s_0-{\scriptstyle 0})=G(s_0+x)$ for $x\leq m$,
so $g_G(x)\leq s_0$ for $x\leq m$, and hence, as $\{x\leq m\}\subset \{g_G(x)\leq s_0\}$,
$S(G)=\inf\{t\ge 0\colon\;P(g_G(x)\leq t)\ge 1/2\}\leq s_0$.
Finally, for $X,Y\sim G$, stoch.\ indep.\, $Q(G)=\inf\{s\colon P(|X-Y|\leq s)\ge 1/4\} \leq s_0$, as
\begin{eqnarray}
P(|X-Y|\leq s_0) &=& \int G(x+s_0)-G(x-s_0-{\scriptstyle 0})\,G(dx)\geq \nonumber\\
&\geq &\int_{[x_0;m]} G(x+s_0)\,G(dx) \geq \int_{[x_0;m]} \frac12\,G(dx)\geq \frac14 \label{eq3}
\end{eqnarray}
Assume $s(G)=s_0$.
In case of kMAD this happens iff $G(m+ks_0)=1/2$, or, equivalently, $G((m;m+ks_0))=0$.
In case of Sn, $S(G)=s_0$ iff $P(g_G(X)> s)  \ge 1/2$ for all $s<s_0$, or, equivalently,
$P(x\colon\; G((x-s_0;x+s_0)) <1/2) \ge 1/2$. But $x-s_0<x_0$ whenever $x< m$, so $G((x-s_0;x+s_0))=G(x+s_0)\ge G(m)=1/2$.
Hence $S(G)=s_0$ iff  $G((x-s_0;x+s_0)) <1/2$ for all $x\ge m$, or, equivalently, iff  $G(x+2s_0 - \scriptsize{0})-G(x) <1/2$ for $x\ge x_0$.
In case of Qn, $S(G)=s_0$ iff the inequalities in \eqref{eq3} are equalities, i.e.; iff $G([x_0;m])=1/2=G(m+s_0)$,
and $\int_{(m;\infty)} G(x+s_0)-G(x-s_0-{\scriptstyle 0})\,G(dx) =0$. The last integral is $0$ iff $G((m;\infty))=0$,
so that altogether, $S(G)=s_0$ iff $G(m)=G(\{\infty\})=1/2$.
\qed\medskip

\end{proof}

\begin{proof}{ to Proposition~\ref{PEprop}: }
For all models, i.e.; GPD, GEVD, Weibull, and Gamma, we can render the scale
estimator arbitrarily large for $\hat Q_3$ sufficiently large, so $\ve_n^\ast\leq  1/4$.
In case of GPD and GEVD, $\hat \beta <0 $ once $\hat Q_3\le 2 \hat Q_2$,
which certainly happens if, in an ideally distributed sample,
we replace all observations $X_i$,
$2\hat Q_2\leq X_i\leq \hat Q_3$ by $\hat Q_2$, entailing \eqref{Nn0def}.
Appealing to Lemma~\ref{HoeffLemm}, up to an event of probability $\LO(\exp(- c n^{\delta}))$ for some $c>0$,
\begin{equation}
\ve_n^\ast = \bar\ve^\ast+\LO_{P^n_\vartheta}(n^{-1/2+\delta/2})
\end{equation}

As \eqref{weibPick} gives valid values for $\xi$ and $\beta$ for any
values of $\hat Q_3$ and $\hat Q_2$, in the Weibull case, we cannot lower the upper bound of $25\%$,
i.e.;  $\lim_n \bar \ve^\ast_n = \bar \ve^\ast=\ve^\ast= 1/4$.\qed
\end{proof}

\begin{proof}{ to Proposition~\ref{MedkMADprop}: }
As we have seen in the considerations in Section~\ref{Sec:MedkMAD}
on producing breakdown, we only can solve (uniquely)
for $\xi$ and $\beta$ as long as the quotient $\hat q_{k;n}$ falls
into $(\check q_k,\bar q_k)$; case-by-case considerations indeed show that
by driving $\hat q_{k,n}$ to either $\check q_k$ (in case of GPD and GEVD)
or $\bar q_k$ (in all cases) produces breakdown, that is,
breakdown could be achieved by either moving all $\hat N_n'$ observations
 from \eqref{Nn1def}
for which $\hat m < X_i\leq \hat m +\hat M_k$ to $(k+1) \hat m$
(entailing $\hat q_{k;n}\approx 1$) or by moving a number of $\hat N_n''$ observations
(as defined in \eqref{Nn2def}) to the interval
 $[(1-\check q_k) \hat m,(k\check q_k+1)\hat m]$ up to the point that it
 contains $n/2$ observations (entailing $\hat q_{k;n}<\check q_k$).
The actual FSBP is then given by the alternative needing to move less observations.
The terms for ABP follow with the usual LLN argument.\qed
\end{proof}

\begin{proof}{ to Proposition~\ref{Nnprop}: }
We start with the fact that for $X_i\iid F$ with Lebesgue density $f$,
the joint c.d.f.\ of the order statistics
$X_{[i_1:n]}$, $X_{[i_2:n]}$ for $1\leq i_1<i_2\leq n$ for $s\leq t$
can be written as
\begin{eqnarray*}
G(s,t)&=&  n \int_{-\infty}^s \!\!\!\!\!f(x)
{\textstyle {n-1 \choose i_1-1}} F(x)^{i_1-1}
\!\!\!\!\! \sum_{k_2=i_2-i_1}^{n-i_1}
\!\!{\textstyle {n-i_1 \choose k_2}} \big(F(t)-F(x)\big)^{k_2}
\bar F(t)^{n-i_1-k_2}\,dx
\end{eqnarray*}
Hence
\begin{eqnarray*}
P(\hat N_n'\ge l) &=&P(X_{[(n/2+l+1):n]} \leq
(k+1) X_{[(n/2+1):n]}) \nonumber\\
&=& n \int_{0}^{1} \!\!\! {\textstyle {n-1 \choose n/2}} u^{n/2} \,
 \sum_{k_2=l}^{n/2-1} {\textstyle {n/2-1 \choose k_2}}
 \big(F(t_k)-u\big)^{k_2} \bar F(t_k)^{n/2-1-k_2}\,du
\end{eqnarray*}
and \eqref{NhatMM} follows by taking differences. Cases
\eqref{NhatPE} and \eqref{NhatkMeM2} follow similarly.\qed
\end{proof}

\begin{proof}{ to Lemma~\ref{HoeffLemm}: }
We note that $\{\hat q_n\leq t\}=\{\sum_i \Jc(X_i\leq t) \ge n \alpha\}$. Hence with
Hoeffding's inequality, \cite{Ho:63}, $P(|\hat q_n-q|\ge t_n) \leq 2 \exp(-2 n (F(t_n+q)-\alpha)^2)$
and (a) follows from $F(t_n+q)-\alpha= f(q) t_n +\Lo (t_n)$.
For (b), note that
$P(\hat I_n \mathop{\scriptstyle \Delta} I) \leq \Ew |F(\hat q_{1,n})-\alpha_1|  + \Ew |F(\hat q_{2,n})-\alpha_2|$.
Hence, for large enough $n$, $P(\hat I_n \mathop{\scriptstyle \Delta} I) \leq 2 f(a_1q_1) |a_1| \Ew |\hat q_{1,n}-q_1| + 2 f(a_2q_2) |a_2| \Ew |\hat q_{2,n}-q_2|$.
and, applying that for a random variable $Z$ taking values in $[0,1]$, for $t\in(0,1)$,
$0\leq \Ew Z \leq t+ \int_{t}^1 P(X>t)$, so by Mill's ratio,
$P(\hat I_n \mathop{\scriptstyle \Delta} I) \leq 2 t + \sum_i {\exp(-2n t^2 f(q_i)^2)}/{(2n t f(q_i)^2)}$.
Plugging in $t=n^{-1/2+\delta}$, we obtain (b).\qed
\end{proof}

\begin{table}[c]
\renewcommand{\arraystretch}{1.7}\addtolength{\tabcolsep}{-1pt}
  \caption{\label{Tab:n40M10000_neg}
Simulated EFSBP  in \% with CLT-based $95\%$-confidence interval (CI) for  $\theta = (\xi=0.7,\beta=1)$;
number of runs is 10000}
\begin{tabular}{p{1.55cm}r p{.75cm}r p{.75cm}r  p{.75cm}r p{0.5cm}r p{.75cm}rp{0.5cm}r p{.75cm}rp{0.5cm}r}
\hline\hline
\multicolumn{1}{l}{Model}
&\multicolumn{1}{c}{Med-}&
&\multicolumn{1}{c}{Med-}&
&\multicolumn{1}{c}{Med-}&
\\[-2.ex]
&\multicolumn{1}{c}{Sn}
&\multicolumn{1}{l}{$\pm$ CI}
&\multicolumn{1}{c}{Qn}
&\multicolumn{1}{l}{$\pm$ CI}
&\multicolumn{1}{c}{\hspace{-1em}$\mbox{kMAD}_{10}$}
&\multicolumn{1}{l}{$\pm$ CI}
&\multicolumn{1}{c}{PE}
&\multicolumn{1}{l}{$\pm$ CI}
\tabularnewline
\hline
GPD $\xi \in \R$&$34.69$&$0.33$&$43.74$&$0.09$&$44.68$&$0.13$&$ 5.94$&$0.10$\tabularnewline
GPD $\xi >0$    &$\hphantom{0}8.78$&$0.18$&$23.44$&$0.21$&$10.65$&$0.07$&$ 5.94$&$0.10$\tabularnewline
GEVD $\xi \in \R$&$ 6.99$&$0.21$&$ 5.89$&$0.21$&$13.38$&$0.24$&$14.85$&$0.13$\tabularnewline
GEVD $\xi >0$    &$ 6.99$&$0.21$&$ 5.89$&$0.21$&$ 4.75$&$0.13$&$7.87$&$0.16$\tabularnewline
Weibull&$37.63$&$0.34$&$40.32$&$0.11$&$47.31$&$0.02$&$25.00^{*}$&$0.00^{*}$\tabularnewline
Gamma&$34.55$&$0.32$&$41.97$&$0.10$&$49.17$&$0.02$&$n.a.$&$-$\tabularnewline
\hline
\end{tabular}
$\hspace*{-1ex} n = 40$
\begin{tabular}{p{1.55cm}r p{.75cm}r p{.75cm}r  p{.75cm}r p{0.5cm}r p{.75cm}rp{0.5cm}r p{.75cm}rp{0.5cm}r}
GPD $\xi \in \R$&$23.55$&$0.21$&$47.51$&$0.04$&$\hphantom{0}44.73$&$0.09$&$ 6.12$&$0.07$\tabularnewline
GPD $\xi >0$    &$12.44$&$0.16$&$18.42$&$0.16$&$\hphantom{0}11.32$&$0.05$&$ 6.12$&$0.07$\tabularnewline
GEVD $\xi \in \R$&$ 3.25$&$0.09$&$ 2.88$&$0.09$&$ 8.86$&$0.14$&$15.01$&$0.09$\tabularnewline
GEVD $\xi >0$    &$ 3.25$&$0.09$&$ 2.88$&$0.09$&$ 6.32$&$0.11$&$6.71$&$0.05$\tabularnewline
Weibull&$26.58$&$0.30$&$45.12$&$0.05$&$47.41$&$0.02$&$25.00^{*}$&$0.00^{*}$\tabularnewline
Gamma&$25.42$&$0.21$&$45.90$&$0.04$&$49.35$&$0.02$&$n.a.$&$-$\tabularnewline
\hline
\end{tabular}
$n = 100$
\begin{tabular}{p{1.55cm}r p{.75cm}r p{.75cm}r  p{.75cm}r p{0.5cm}r p{.75cm}rp{0.5cm}r p{.75cm}rp{0.5cm}r}
GPD $\xi \in \R$&$21.86$&$0.03$&$49.75$&$0.00$&$\hphantom{0}44.75$&$0.03$&$ 6.38$&$0.03$\tabularnewline
GPD $\xi >0$    &$14.99$&$0.13$&$16.06$&$0.02$&$\hphantom{0}11.82$&$0.02$&$ 6.37$&$0.03$\tabularnewline
GEVD $\xi \in \R$&$ 1.06$&$0.03$&$ 1.27$&$0.03$&$7.25$&$0.05$&$15.39$&$0.04$\tabularnewline
GEVD $\xi >0$    &$ 1.06$&$0.03$&$ 1.27$&$0.03$&$7.22$&$0.05$&$6.20$&$0.08$\tabularnewline
Weibull&$19.77$&$0.03$&$49.01$&$0.01$&$47.55$&$0.01$&$25.00^{*}$&$0.00^{*}$\tabularnewline
Gamma&$24.13$&$0.04$&$49.16$&$0.01$&$49.46$&$0.01$&$n.a.$&$-$\tabularnewline
\hline
\end{tabular}
$\hspace{1em}\vspace*{4ex} n \hfill 
= \hfill 
1000$

$*$ : theoretical values,\\
$n.a.$: not available; in these cases, $25 \%$ is an upper bound
\end{table}
\section*{Acknowledgement}
We thank two anonymous referees for their helpful comments.

\begin{figure}[l]
\includegraphics[scale=0.7]{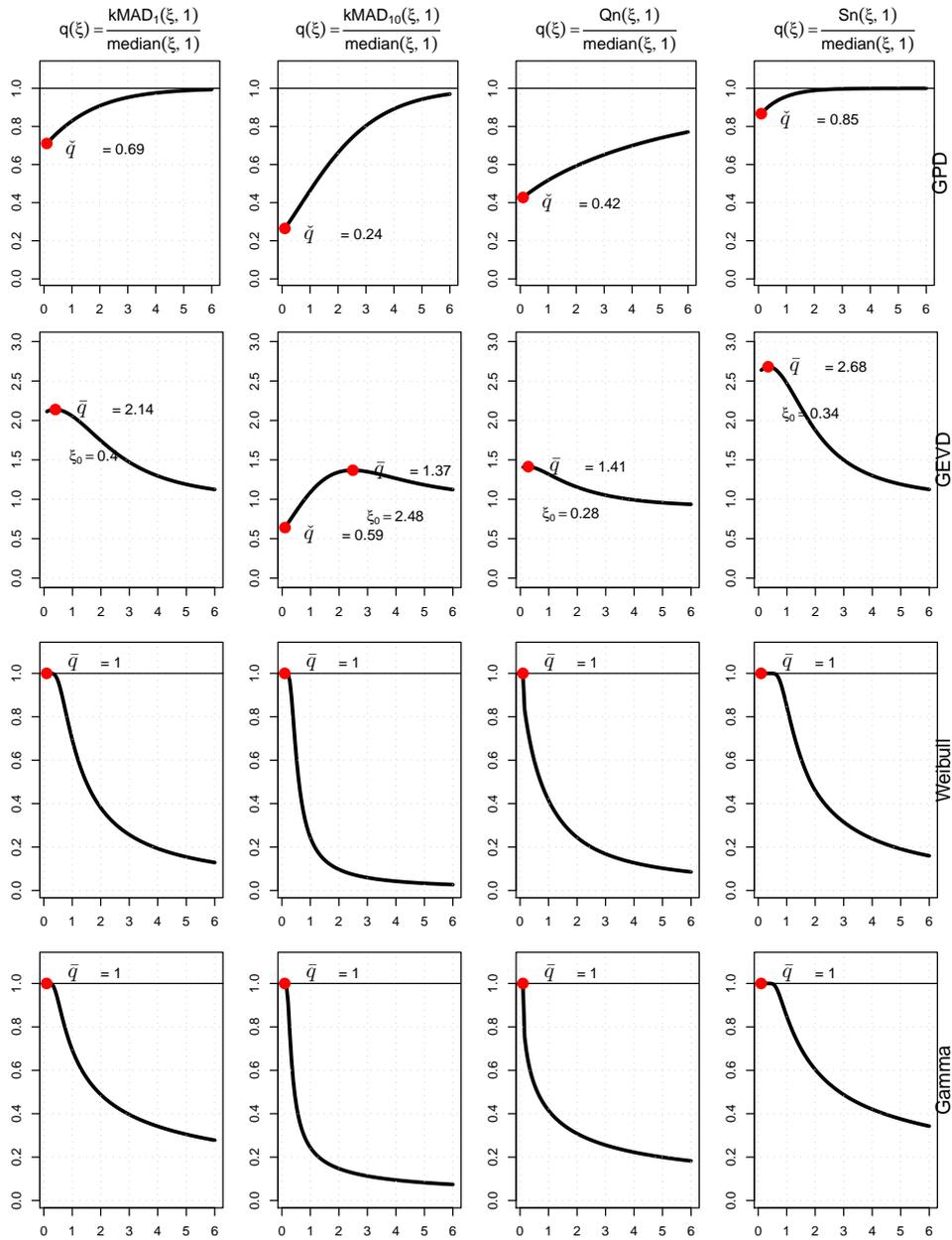}
%
 \caption{\label{fig:LD_quotients} Quotients $\rm kMAD(\xi,k=1)/\rm med(\xi)$ and $\rm kMAD(\xi,k=10)/\rm med(\xi)$,
$\rm Qn(\xi)/\rm med(\xi)$ and $\rm Sn(\xi)/\rm med(\xi)$ as functions in $\xi$; we also
include with respective $\check q$, $\bar q$}
\end{figure}



\begin{small}

\end{small}
\end{document}